\documentclass[superscriptaddress,notitlepage,preprint]{revtex4-1}
\usepackage[utf8]{inputenc}
\usepackage[T1]{fontenc}
\usepackage{bm}
\usepackage{graphicx}
\usepackage{xcolor}
\usepackage{amsmath}
\usepackage{amssymb}
\usepackage{siunitx}
\usepackage{parskip}

\begin{document}

\title{Molecular chirality controls droplet division and helical fiber formation in liquid crystal emulsions}
\author{Simon Čopar}
\affiliation{Faculty of Mathematics and Physics, University of Ljubljana, Jadranska cesta, 1000 Ljubljana, Slovenia.}
\author{Mariana V. M. Rodriguez}
\affiliation{Condensed Matter Physics Department, Jožef Stefan Institute, Jamova 39, 1000 Ljubljana, Slovenia.}
\author{Peter Marinko}
\affiliation{Faculty of Mathematics and Physics, University of Ljubljana, Jadranska cesta, 1000 Ljubljana, Slovenia.}
\author{Seyed Reza Seyednejad}
\affiliation{Faculty of Mathematics and Physics, University of Ljubljana, Jadranska cesta, 1000 Ljubljana, Slovenia.}
\author{Jan Dolinar}
\affiliation{Faculty of Mathematics and Physics, University of Ljubljana, Jadranska cesta, 1000 Ljubljana, Slovenia.}
\author{Miha Škarabot}
\affiliation{Condensed Matter Physics Department, Jožef Stefan Institute, Jamova 39, 1000 Ljubljana, Slovenia.}
\author{Karthik Peddireddy}
\affiliation{Dynamics of Complex Fluids, Max Planck Institute for Dynamics and Self-Organization, Am Fassberg 17, 37077 Göttingen, Germany.}
\author{Khoa V. Le}
\affiliation{Department of Chemistry, Faculty of Science, Tokyo University of Science, 1-3 Kagurazaka, Shinjuku-ku, Tokyo 162-8601, Japan.}
\author{Samo Kralj}
\affiliation{Condensed Matter Physics Department, Jožef Stefan Institute, Jamova 39, 1000 Ljubljana, Slovenia.}
\author{Slobodan Žumer}
\affiliation{Faculty of Mathematics and Physics, University of Ljubljana, Jadranska cesta, 1000 Ljubljana, Slovenia.}
\author{Miha Ravnik}
\affiliation{Faculty of Mathematics and Physics, University of Ljubljana, Jadranska cesta, 1000 Ljubljana, Slovenia.}
\affiliation{Condensed Matter Physics Department, Jožef Stefan Institute, Jamova 39, 1000 Ljubljana, Slovenia.}
\author{Venkata S. R. Jampani}
\email{vsrao.jampani@ijs.si}
\affiliation{Condensed Matter Physics Department, Jožef Stefan Institute, Jamova 39, 1000 Ljubljana, Slovenia.}

\begin{abstract}
Molecular chirality is a source of broken mirror symmetry, but using it to control mesoscale structures with a tunable length scale remains challenging. 
Here, we demonstrate that adding a chiral dopant to nematic liquid crystal droplets 
bounded by a deformable two-surfactant interface controls their morphogenesis: the ratio of droplet diameter to cholesteric pitch determines whether droplets divide asymmetrically or symmetrically upon cooling, and whether they transform into single- or double-strand helical fibers. The fiber periodicity and thickness both scale linearly with the cholesteric pitch, which varies by less than $2\,\%$ with temperature across the self-shaping window. Numerical simulations reveal that chirality-driven elastic stresses at the interface destabilize the droplets and trigger cusp-mediated shape transformations. These results establish cholesteric pitch as a design variable to precisely control droplet division and decouple the dimensions of spontaneously formed mesoscale structures from temperature dependence.
\end{abstract}

\maketitle

\section{Introduction}
The term chirality, coined by Lord Kelvin, describes the non-superposability of an object on its mirror image, as illustrated by human hands~\cite{kelvin1894molecular}. Molecular chirality is central to the chemistry of life, and the origin of homochirality is a persistent question across scientific disciplines \cite{mason1984origins,blackmond2010origin}. 
At the molecular level, chirality can initiate symmetry-breaking processes that lead to handedness of self-organized structures and the directionality of motion, such as hydrogen-bonded polypeptide helices and the DNA double helix \cite{pauling1951structure, watson1953molecular}, and unidirectional rotation of light-driven molecular motors \cite{koumura1999light}. In living systems, such chiral asymmetries can propagate across scales and influence morphogenesis at the tissue level \cite{LiZ_PhysRevLett132_2024}. In passive ordered liquids such as liquid crystals (LCs), chirality was first discovered in biologically derived cholesterol compounds, which give chiral nematics their common name ``cholesterics''. In these phases, molecular chirality imposes a spontaneous helical twist of the average molecular orientation, which is quantified by an additional length scale called the cholesteric pitch -- the distance corresponding to a full $360^\circ$ rotation of molecules \cite{deGennes, harris1999molecular}. The rich variety of non-uniform orientational textures that form in the bulk of chiral nematics, ranging from fingerprint textures, torons, merons, concentric shells, and more, has led to many fundamental results and applications \cite{TaiJB_LiquidCrystalsToday32_2023}. In contrast to these bulk orientational textures, whether an intrinsic chiral length scale can quantitatively organize the shape transformations of a freely deforming interface remains unexplored. Chirality is known to reorganize the internal textures of liquid crystal droplets \cite{orlova2015creation,SecD_NatCommun5_2014}, but its role in defining a free-interface morphogenetic pathway and dimensions has not been established. Here, we exploit the cholesteric pitch -- chemically tunable, temperature-stable length scale -- as a control parameter that programs chiral nematic emulsion droplets into mesoscale structures with defined dimensions.

Liquid crystalline order assists in reshaping droplets themselves. More than four decades ago, Lavrentovich and Nastishin reported spontaneous division of cholesteric liquid crystal (CLC) droplets near the cholesteric-smectic-A phase transition, which is accompanied by a transformation from spherical droplets to cylindrical fibers in a surfactant-containing glycerol medium~\cite{lavrentovich1984division}. Later studies showed related cylindrical fiber morphologies in binary liquid-crystal mixtures that form the smectic A phase on cooling from the isotropic phase~\cite{arora1989reentrant, pratibha1992cylindrical, adamczyk1995nematoid, naito1997pattern, kim2019self, browne2025structural}. More recent works extended droplet morphogenesis to include cylinder-to-disc-like and cylinder-to-coil transformations~\cite{browne2026arrested,sato2026morphological}. Similarly, in the nematic phase, droplet-to-cylindrical fiber transitions have also been achieved. Nevertheless, these approaches have so far relied on specific molecular designs, such as mesogens covalently linked to surfactant head groups~\cite{toquer2008colloidal}, heterogeneous nematic oligomers that form branched fiber networks~\cite{wei2019molecular}, or two-surfactant strategies that balance bulk elasticity and interfacial tension~\cite{PeddireddyK_ProcNatlAcadSci118_2021}. Separately, achiral bent-core LCs were shown to exhibit spontaneous symmetry breaking into racemic chiral bulk states~\cite{jakli2018physics, link1997spontaneous} and helical nanofilament structures~\cite{hough2009helical}, with chiral dopants providing control over handedness and enantiomeric excess~\cite{thisayukta2000distinct}. However, helical structures with predefined pitch formed directly from droplets of simple rod-shaped liquid crystals, together with pitch-controlled droplet divisions, have not been demonstrated.

Here, we show that thermotropic CLC droplets dispersed in aqueous surfactant solutions undergo pitch-programmed self-shaping. Cooling modifies the surfactant-mediated interfacial tension, which leads to shape instability, while the cholesteric pitch remains nearly constant throughout the shape morphogenesis temperature window ($<2\,\%$ variation). Temperature cycling, therefore, acts as the trigger for driving shape transformation, whereas the ratio of the initial droplet diameter to the cholesteric pitch organizes irreversible asymmetric and symmetric droplet divisions, as well as reversible single- and double-strand helical fibers formation. Within the experimentally explored range, we demonstrate that the cholesteric pitch determines the characteristic dimensions of the resulting structures.

\section{Results}
We used the nematic LC material E7 as the host medium and induced chirality by adding a chiral dopant S811. Dopant concentrations 0.1 to 15 wt\% were used to tune the cholesteric pitch over a broad range while maintaining cholesteric phase across the experimental temperature window \cite{chen2004phase}. For each composition, the pitch was measured using the Cano wedge method \cite{smalyukh2002three}, and also independently estimated from the concentric layered structure formed in homeotropically anchored droplets \cite{orlova2015creation, SecD_NatCommun5_2014}. For each desired pitch, the CLC mixture is then mixed with a fixed 2 wt\% non-ionic surfactant (monoolein), after which the chiral droplets are dispersed in an aqueous CTAB surfactant solution at a concentration well below the critical micelle concentration. The self-shaping is achieved with the two-surfactant approach, with temperature acting as the tuning parameter for interfacial tension, and thus driving the shape morphology \cite{PeddireddyK_ProcNatlAcadSci118_2021}.

\begin{figure}
    \centering
    \includegraphics[width=\linewidth]{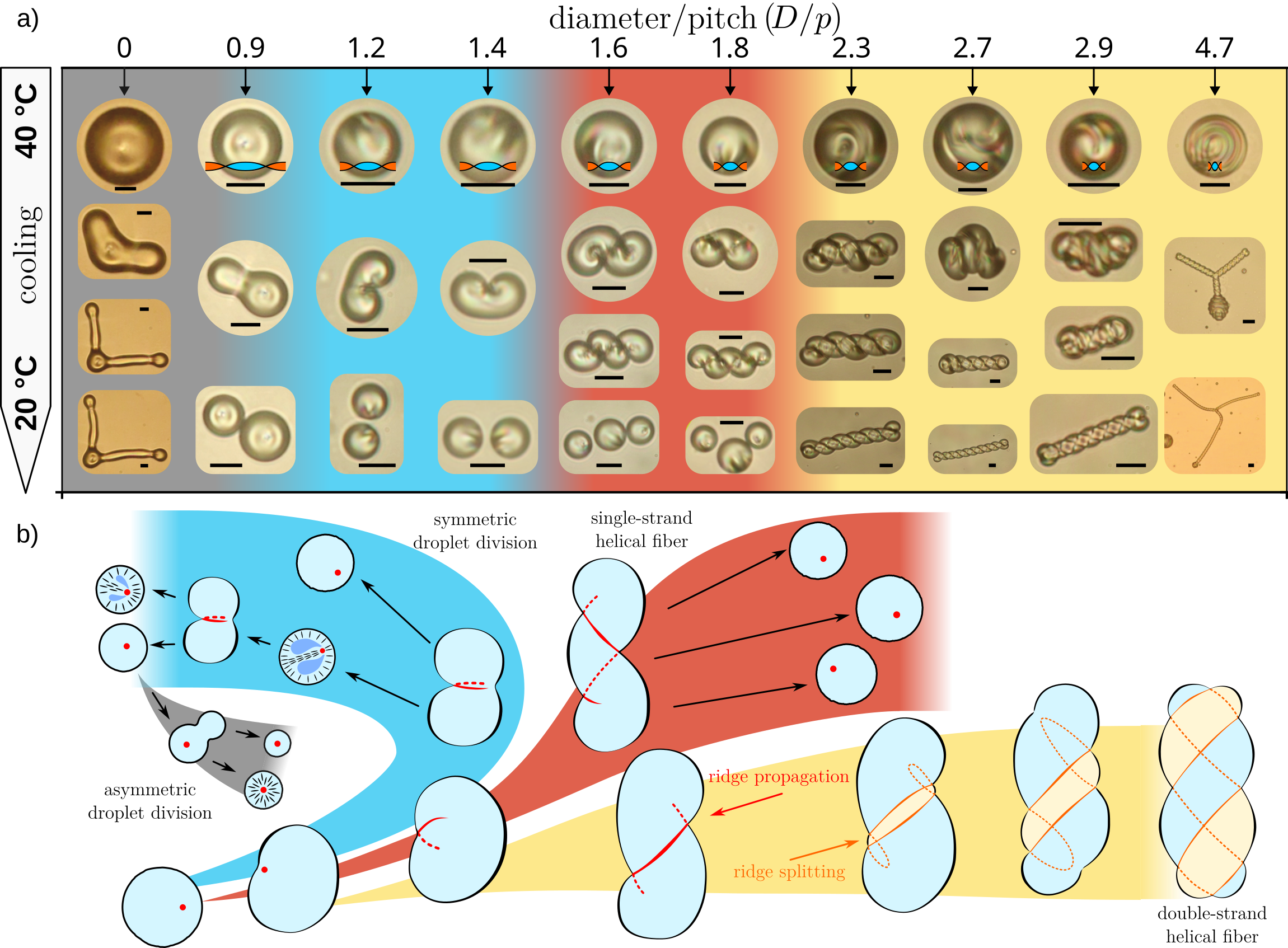}
    \caption{
    \textbf{Droplet morphogenesis with temperature, as a function of diameter-to-pitch ratio, $D/p$}. (a) Self-shaping droplet morphogenesis sequences upon cooling (vertical axis). Increasing $D/p$, either by increasing the initial droplet diameter or by decreasing the cholesteric pitch, produces a progression from the achiral nematic limit, where a droplet grows into cylindrical fibers (two, in this case), through asymmetric droplet division at very low chirality (grey shaded region), symmetric cusp-mediated division (blue region), single-strand helical fibers (red region), and double-strand helical fibers (yellow region). At the highest $D/p$ values, fiber branching is also observed. Scale bars: \SI{10}{\micro m}. Bow-tie diagrams show the full $2\pi$ intrinsic cholesteric pitch of each mixture. (b) Illustration of the morphological pathways, with background color matching (a). In three droplets, the director field is shown schematically, from completely radial in small droplets, to an increasingly large cholesteric bubble with out-of-plane director (blue shading), accompanying an off-center defect. The concentric layered director structure of the largest droplets is not depicted schematically, but it is visible in the experimental image  of the initial droplet in panel (a) at $D/p=4.7$.
    }
    \label{fig:phase_diagram}
\end{figure}

Upon cooling, the droplets divide or transform, as summarized in the morphology map shown in Fig. \ref{fig:phase_diagram}. The principal organizing parameter is the diameter-to-pitch ratio, $D/p$, where $D$ is the initial droplet diameter and $p$ is the intrinsic cholesteric pitch. This ratio organizes the observed morphologies into distinct regimes: asymmetric droplet divisions, symmetric droplet divisions, single-strand helical fibers, and double-strand helical fibers. Throughout this work, $D/p = 0$ denotes the achiral limit ($p\rightarrow\infty$).
With shorter pitches $p$, and thus larger $D/p$, the morphology deviates from both conventional droplet-to-cylinder transformations \cite{arora1989reentrant,pratibha1992cylindrical,adamczyk1995nematoid,naito1997pattern} and previously reported CLC droplet division near the cholesteric–smectic-A transition \cite{lavrentovich1984division}. All transformations that do not involve droplet division are reversible under a narrow window of temperature cycling, typically between 40 $^\circ$C and 30 $^\circ$C, whose position can be adjusted through the surfactant coverage. Within this window, the pitch shows no measurable change by the Cano wedge method and varies by less than 2\% as measured from droplet textures.

First, we consider the smallest $D/p$ values, which correspond to approximately an achiral nematic limit. In the pure achiral nematic phase, where the pitch is effectively infinite, droplets contain a point defect at the center, with a completely symmetric radial director field. Upon cooling, the droplets grow into straight cylindrical fibers as shown in $D/p=0$ \cite{PeddireddyK_ProcNatlAcadSci118_2021}. At low but nonzero chiralities, for $D/p<1$, the defect remains approximately centered in the droplet.  Under these conditions, the droplet still starts growing a cylindrical fiber, as in the achiral limit, but it only grows into a short rounded stub, resulting in an asymmetric dumbbell-like shape and then divides into two daughter droplets of unequal size on cooling (grey region in Fig. \ref{fig:phase_diagram}). 

\begin{figure}
    \centering
    \includegraphics[width=\linewidth]{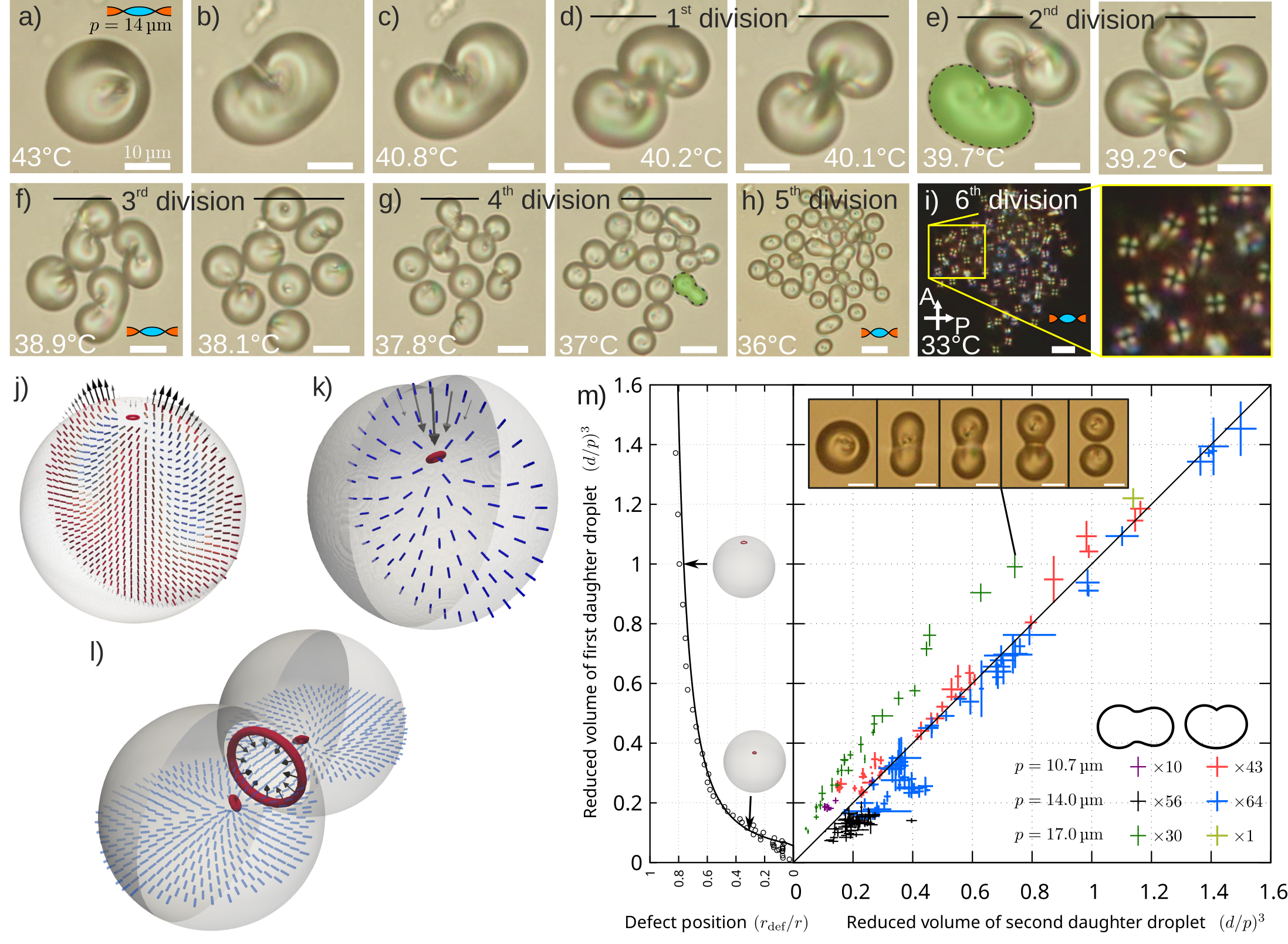}
    \caption{
    \textbf{Droplet division sequence.}
    (a-d) Upon cooling, a droplet divides through a cusp-shaped deformation into two identical daughter droplets. (e-g) The second to fourth divisions proceed in the same way. Cusp shape is shaded in green in (e). (h-i) Sufficiently small droplets instead deform into a dumbbell shape and divide unevenly. The dumbbell outline leading to the 5th and 6th division is marked in (g). At this stage, the defect returns to the center of the droplet, and symmetry is restored, as confirmed by the crossed polarizers image (i), which resembles an achiral nematic droplet. (j-l) Simulated director textures at successive stages of division. Arrows at the surface indicate the normal elastic forces that drive the reshaping. The surface is pulled inward near the defect and pushed outwards in the neighboring regions on either side, producing a cusp, which develops into a pair of $+1$ point defects and a $-1$ loop defect at the neck before division. (m) Daughter droplet volumes after division, with experimental error margins obtained by measuring the diameters across several frames of the video (Movie S2). Size order of droplets is chosen to visually separate different experiments above and below the symmetry line. For the lower two pitches, cusp-mediated divisions yield nearly equal daughter volumes, and dumbbell-mediated divisions more often result in unequal volumes. At the largest pitch, almost all droplets divided unevenly (inset). The left subpanel shows the simulated radial defect offset from the center relative to the droplet radius, for a gradual increase of pitch, and a rational function fit as a visual guide.
    The legend shows the number of divisions measured for each case.
    }
    \label{fig:division}
\end{figure}

For symmetric division,  the prerequisite is to have an eccentric defect. For $D/p$ between $1$ and $\sim 1.5$, chirality breaks the radial symmetry by pushing the point defect away from the droplet center, resulting in a cholesteric bubble structure, as defined in Posnjak et al.~\cite{PosnjakG_NatCommun8_2017} and also shown in Figure \ref{fig:division}j with blue rods showing out-of-plane director. Numerically simulated defect displacement as a function of relative droplet size is shown in the left part of Figure \ref{fig:division}m. Upon cooling, a cusp forms on the surface closest to the eccentric defect. This defect-interface coupling generates anisotropic inward stress at the LC-water interface, producing localized inward curvature similar to a cusp of an apple \cite{chakrabarti2021cusp}. The closer the defect lies to the surface, the greater the maximum curvature of the cusp. As cooling proceeds, the droplet folds at the cusp into a bean shape and the deformation wraps around the droplet, eventually forming a neck. The droplet then divides into two daughter droplets of nearly equal volume (blue region in Fig.~\ref{fig:phase_diagram}).

When droplets are large enough relative to the pitch, instead of a simple cholesteric bubble, layered cholesteric structures form inside the droplet \cite{orlova2015creation, SecD_NatCommun5_2014}, and cooling drives fiber formation instead of division. 
For $D/p$ between $\sim 1.5$ and $\sim 2$, droplets transform into single-strand helical fibers. In this regime, the localized cusp does not close into a division neck; instead, the initial cusp extends into a long ridge that winds helically around the growing fiber and subsequently divides into several (typically 3) unequal droplets (red region in Fig.~\ref{fig:phase_diagram}). For $D/p > 2$, droplets contain multiple director windings and preferentially transform into double-strand helical fibers -- a single connected liquid volume bearing two helically winding surface ridges (yellow region in Fig.~\ref{fig:phase_diagram}). 
In the lower part of this regime, the transformation starts with single-strand fiber, but the ridge becomes unstable and splits, forming a continuous subsurface half-integer disclination loop along the fiber, resulting in a double-strand form. For sufficiently large initial droplets, such a disclination loop is present from the start, similar to structures reported in simulations by Seč et al.~\cite{SecD_NatCommun5_2014}, and the fiber grows directly from the region where the disclination loop turns around, until the entire droplet is converted into the double-strand fiber. The fiber diameter is set by the cholesteric pitch, while its length is set by the initial droplet volume. The larger droplets sometimes lead to branched fibers ($D/p=4.7$ in Fig.~\ref{fig:phase_diagram} and Movie S1).

In the symmetric-division regime, droplets undergo successive equal volume divisions with cooling, while the cholesteric pitch stays effectively fixed (Fig.~\ref{fig:division}a-g). Each division reduces the daughter droplet diameter to $\approx 79\,\%$ of the parent diameter, so after three such successive divisions, the droplet diameter is reduced by roughly a factor of two, which is comparable to the width of the symmetric division window in the morphology map (Fig.~\ref{fig:phase_diagram}). Further divisions are therefore asymmetric, and proceed via dumbbell-deformation, as the droplets become too small relative to the pitch and the chirality-induced symmetry breaking becomes weaker (see Fig.~\ref{fig:division}h-i).

To elucidate the mechanism of division, we simulated a spherical droplet containing an eccentric point defect associated with a cholesteric bubble structure (Fig.~\ref{fig:division}j). From the simulated director field, we computed the elastic stress tensor and extracted normal forces acting on the droplet surface (black arrows). These forces pull the surface inwards above the near-interface defect while pushing the neighboring surface regions outwards, consistent with the location and orientation of the observed cusp (Fig.~\ref{fig:division}k). The cusp then elongates and extends around the entire droplet, forming a neck accompanied by a newly formed disclination ring, which is topologically compensated by a new point defect. In the state just before division, each lobe already contains its own point defect, and the disclination loop at the neck tightens and pinches the droplet into two daughters (Fig.~\ref{fig:division}l). Note that this process is passive in the sense that it neither requires chemical fuel nor internally generated active stress, in contrast to cytokinesis in living cells, in which the actomyosin contractile ring forms through protein self-assembly and actively drives symmetric cell division~\cite{green2012cytokinesis}. Related division-like deformations have also been reported in active nematic or chemically active droplets, where active stresses, defect structure, membrane elasticity, and surface tension together shape the interface morphologies~\cite{zwicker2017growth, giomi2014spontaneous, leoni2017defect}.

Quantifying volumes of daughter droplets numerically from video after each division for 3 different cholesteric pitches, as shown in Figure~\ref{fig:division}m, confirms that cusp-mediated divisions result in daughter droplets of approximately equal volume and dumbbell-mediated divisions in unequal volumes of daughter droplets. The left sub-panel of Figure~\ref{fig:division}m shows the simulated defect position relative to the droplet radius, indicating where the defect resides when each type of division takes place. At the longest pitch studied ($p=17\,{\rm \mu m}$), droplets rarely form cusp-driven symmetric shapes, suggesting that at larger diameters the same \emph{relative} defect offset places the defect farther from the surface, weakening the localized deformation and making it harder to generate an initial cusp. Thus shifting the division toward the dumbbell-mediated regime.

\begin{figure}
    \centering
    \includegraphics[width=1.\linewidth]{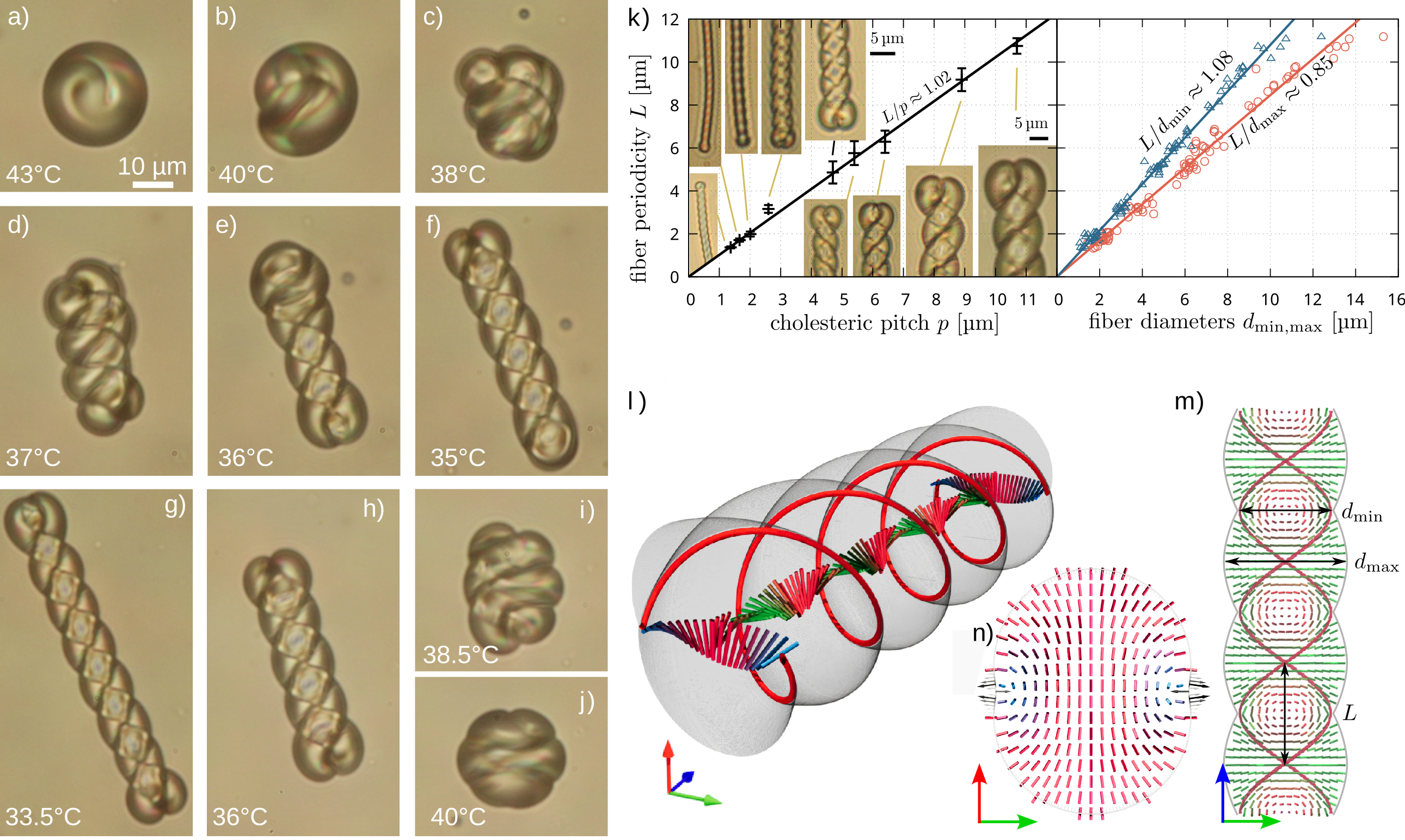}
    \caption{
    \textbf{Formation of double-strand helical fibers from cholesteric droplets.} (a-j) Reversible cooling and heating show the transformation of a surface cusp into a ridge, which then develops into a double-strand fiber and back into a droplet. (k) The fiber periodicity $L$ and the characteristic diameters $d_{\text{min}}$ and $d_\text{max}$ scale linearly with the pitch, showing that cholesteric pitch is the only relevant length scale. (l-n) Projections of the numerical simulations show how the double-strand fiber accommodates twist across the diameter from disclination to disclination, and along its length. Panel (m) defines geometrical quantities plotted in (k), and nearly circular double-twist cores are visible in the top view through each ``diamond'' of the fiber.
    }
    \label{fig:fiber}
\end{figure}

Figure \ref{fig:fiber} shows a representative droplet at $D/p \approx 2.4$ that undergoes a gradual, controlled transformation into a double-strand fiber on cooling to $33.5\,{}^\circ{\rm C}$. Within this thermal-cycling window, the transformation is fully reversible (Fig. \ref{fig:fiber}a-j and Movie S3). Double-strand fibers form across dopant concentrations from 1 to 15 wt\%. The key condition is that the initial droplet is sufficiently large to favor disclination loops over a single eccentric point defect. Under the interfacial conditions studied, the cholesteric pitch sets the characteristic transverse and axial dimensions, whereas volume conservation implies that the initial droplet volume primarily determines the fiber length. The repeat length $L$ (the axial distance over which the double-strand morphology repeats) and the diameters $d_\text{min}$ and $d_\text{max}$ all scale linearly with the pitch $p$ over the measured range (Fig.~\ref{fig:fiber}k), while the pitch itself follows the expected inverse dependence on dopant concentration. Therefore, by choosing the pitch, we can precisely predetermine fiber dimensions. In particular, the fiber length increases directly proportionally to $p$ and cross-sectional area scales as $p^2$. For an initial droplet of volume $v_0$, the overall fiber length is expected to scale roughly as $v_0/p^2$.

We complemented the experiments with numerical modeling in a fixed geometry matched to the experiment. In the strong anchoring limit, the anchoring contribution is treated as a fixed boundary condition, and a surface term encapsulates all the interactions between the LC, water, and both surfactants at the interface. The simulated director profile reproduces the confocal observations of two opposing cusps, each associated with the sub-surface disclination line, resulting in a nephroid-like cross-section in the plane perpendicular to the fiber axis (Fig. \ref{fig:fiber}n). This cross-section remains geometrically the same along the entire fiber but rotates helically. Consequently, the longitudinal cross-section alternates between $d_\text{min}$, measured between cusps, and $d_\text{max}$, measured after $90^{\circ}$ rotation (Fig. \ref{fig:fiber}m). The twist is distributed between the longitudinal and radial directions instead of localizing all twist along a single direction (Fig.~\ref{fig:fiber}l), and approximately constant ratio $L/d_\text{min}\approx 1$ indicates that the double-strand fiber maintains a fixed geometric relation between axial and transverse twist over the full pitch range. Upon further cooling beyond the reversible window, the fiber destabilizes into individual droplets of diameter close to $d_\text{max}$. Such a destabilization is consistent with Fig.\ref{fig:fiber}k, the resulting droplets fall within the symmetric division regime $D/p < 1.5$ (Movie S5), returning the system to the division regime of Fig.~\ref{fig:phase_diagram}. Temperature, therefore, determines when the instability is activated, $D/p$ selects the morphological regime, and the cholesteric pitch fixes the characteristic dimensions.

\section{Discussion}

The droplet morphologies described here require an interface that can deform under elastic stress~\cite{wei2019molecular, yuan2024intrinsic}. In previously reported self-shaping systems, the transformation was linked to an interfacial layer formed by surfactants added to either the LC or the aqueous phase during cooling from the isotropic phase~\cite{bahr2006surfactant, bahr2009experimental}. Similar interfacial layers drive shape transformations of isotropic alkanes droplets in aqueous surfactant solutions, confirming that a nanoscale interfacial layer can drive shape transformation upon cooling~\cite{guttman2016faceted, guttman2019nanostructures, denkov2015self}. Additionally, a morphological transition from a spherical to a crumpled state has been reported in lipid vesicles, driven by coupling between lipid orientational order, curvature, and topological defects \cite{hirst2013morphology}. In the present system, monoolein in the LC phase and CTAB in the aqueous phase are expected to promote excess interfacial order during cooling, as in the earlier two-surfactant self-shaping mechanism\cite{PeddireddyK_ProcNatlAcadSci118_2021} while coupled additionally to the eccentric defect. Recent simulations further support this mechanism, showing that surfactant combinations at oil–water interfaces can generate surface layers that reduce the effective interfacial tension to very low values \cite{hendrikse2026quantitative}. The symmetric-division and helical-fiber pathways begin with a localized cusp or ridge(Figs. \ref{fig:division}k, and \ref{fig:fiber}l-n). Their spontaneous increase in interfacial area is consistent with effectively negative temperature-controlled interfacial tension of our two-surfactant system \cite{PeddireddyK_ProcNatlAcadSci118_2021}. In isotropic liquids, negative interfacial tension is unstable and normally leads to spontaneous emulsification. In liquid crystals, however, bulk elasticity opposes the tendency to increase surface area, thereby arresting the uncontrolled emulsification, allowing finite, temperature-controlled shape transformations to emerge. The shape transformation is triggered when the free energy for the same volume of LC is lower in a nonspherical or divided morphology than in an initial spherical droplet. 

In a non-chiral system, the characteristic length scale is given by $r_{\gamma}=-3K/\gamma$, the equilibrium radius of the extending cylindrical fibers, such as those seen at $D/p=0$ in Fig.~\ref{fig:phase_diagram}a. Here, $K$ is the elastic constant and $\gamma$ the effective interfacial tension, which must be effectively negative to stabilize a nonspherical shape. This length scale depends strongly on temperature within the self-shaping window of 10 $^\circ$C and is not suitable for producing well-controlled dimensions \cite{PeddireddyK_ProcNatlAcadSci118_2021}. Molecular chirality introduces a second length scale, the cholesteric pitch, $2\pi/q_0$, which changes by <2\% within the same self-shaping temperature window. Our results show that pitch is the dominant scale that sets the fiber dimensions, whereas competition with interfacial-elastic scale controls the transition between division and helical fiber regimes. Unlike in non-chiral systems, where fiber dimensions are sensitive to the temperature dependence of $r_{\gamma}$~\cite{PeddireddyK_ProcNatlAcadSci118_2021}, the chiral system exhibits stable and reproducible fiber dimensions over a wide range of dopant concentrations. Although we cannot measure $\gamma$ independently, $r_{\gamma}$ is estimated from the straight cylindrical fiber thickness in the non-chiral limit (Fig.~\ref{fig:phase_diagram} at $D/p=0$). Assuming that $\gamma$ and $K$ do not dramatically change with the chiral dopant concentration, we can qualitatively compare this length scale with the double-strand helical fiber diameter in the chiral regime (Fig.~\ref{fig:phase_diagram} at $D/p=2.3$--$4.7$), and observe that they are of the same order of magnitude. This suggests that the observed morphological diversity arises from the competition between two length scales, one set by the tension-elasticity balance and the other by the cholesteric pitch, with different morphogenesis regimes emerging as their ratio varies.

The double-strand helical fiber structure should be regarded as a single continuous volume of liquid, even though it resembles a tightly coiled pair of independent fibers. Double-twist-like fibers have been reported in lyotropic systems, where lipid molecules assemble into smectic-A layers in a cylindrical geometry, known as myelin structures \cite{SakuraiI_MolCrystLiqCryst180_1990}. Unlike ours, in these systems, fibers grow through layer extension and folding, and double-twisted morphologies arise geometrically from the coiling of two separate fibers, similar to the tightly packed solid ropes \cite{PrzybyS_EurPhysJE4_2001}. In contrast, the double-strand fibers reported here form through chirality-driven cusp and ridge formation within a single continuous CLC volume, with the disclination lines acting as anchors for the ridges, similar to soap films stretched between helically shaped wires \cite{MayH_ProcRSocA464_2008}.

The presented system is entirely passive, demonstrating that localized elastic stresses coupled to chirality-driven topological defects are sufficient to precisely induce three-dimensional shape transformations that, after formation, remain stable. In contrast, in active soft matter, defect-guided morphogenesis is achieved through continuous energy input, and results in dynamically changing shapes until energy is depleted \cite{keber2014topology,metselaar2019topology}.

\section{Conclusion}
Results presented here show that cholesteric pitch, when coupled to a deformable surfactant-mediated interface, acts as a tunable mesoscale parameter that dictates controlled division and the growth of single- and double-strand helical fibers. Because this length scale is predictable and nearly temperature-independent, it allows straightforward control over the dimensions of the shaped structures, while letting temperature independently trigger shape transformation through changes in interfacial tension. Our system is minimal, relying solely on commercially available thermotropic calamitic liquid crystals, chiral dopants, and surfactants, yet it reproduces geometrical shapes associated with living matter, such as contractile-ring-like division, helical filament formation, and defect-mediated stress localization~\cite{ravichandran2025topology, maroudas2021topological}, entirely passively. The phenomena are not specific to the selected materials, and other combinations of chiral liquid crystals with ionic and non-ionic surfactants should lead to similar behavior, with shape differences mostly depending on the intrinsic elastic properties of the chosen material. In this abiotic chiral droplet system, the controlling length scale is the cholesteric pitch, which utilizes molecular chirality to achieve a programmable mesoscale structure. The observed relation between helical fiber pitch, diameter, and cholesteric pitch suggests a route to monodisperse chiral soft-matter architectures with predefined pitch, diameter, and handedness. Ultimately, this discovery allows us to design soft materials for the bulk production of uniformly shaped microfibers at an industrial scale.

\subsection{Numerical modeling}

Numerical results of the different stages of the droplet division process are obtained by minimising the Landau--de Gennes free energy volume density
$f_{LdG}=\frac{A}{2} \mathrm{Tr}Q^2 + \frac{B}{3}\mathrm{Tr}Q^3 + \frac{C}{4}\left(\mathrm{Tr}Q^2\right)^2 + \frac{L_{1}}{2}\partial_k Q_{ij}\partial_k Q_{ji} + 2 L_{1} q_{0}\, \epsilon_{ijk} Q_{il} \partial_k Q_{jl}$
in Q-tensor form~\cite{deGennes}, integrated across the droplet volume.
The Q tensor describes the local order of the nematic liquid crystal because the major eigenvalue and eigenvector provide the local scalar order parameter and the nematic director, respectively.
$A$, $B$ and $C$ are material dependent coefficients, $L_{1}$ is the single elastic constant (related to the Frank constant by $K_{F}=\tfrac{9}{2}S_{0}^{2}L_{1}$), and $q_{0}=2\pi/p$ is the intrinsic chiral wavevector of the cholesteric LC, with $p$ the cholesteric pitch.
All the parameter quantities are taken from refs \cite{stark2001physics, kralj1991nematic}, giving $A=-1.72\times 10^{5}\,\mathrm{J/m^{3}}$, $B=-2.12\times 10^{6}\,\mathrm{J/m^{3}}$, $C=1.73\times 10^{6}\,\mathrm{J/m^{3}}$, equilibrium uniaxial order $S_{0}=0.533$, and $L_{1}=1.6\times 10^{-11}\,\mathrm{N}$ (corresponding to a nematic correlation length $\xi^2=L_{1}/(A+BS_{0}+\tfrac{9}{2}CS_{0}^{2})$ giving $\xi=10\,\mathrm{nm}$, rescaled compared to experiment to allow for use of lower resolution numerical grid).
Homeotropic anchoring is imposed on the interface by a Nobili--Durand surface potential 
$f_{ND}=\frac{W}{2}\mathrm{Tr}\left(Q-Q^{0}\right)^2$,
where $Q^{0}$ is determined by the surface normal. The anchoring strength is chosen as $W=10^{-2}\,\mathrm{J/m^{2}}$.
We employ a finite element method with a conjugate gradient algorithm to minimize the total free energy, using linear basis functions within tetrahedral mesh elements.
The force density at the interface is calculated by multiplying the normal surface vector by the nematic stress tensor, ${\rm d}F_i=\sigma_{ij}\nu_j {\rm d}S$, in the equilibrium state,
with $\sigma_{ij}=\delta_{ij}(f+P)-\frac{\delta f}{\delta(\partial_j Q_{kl})}\partial_iQ_{lk}$, where $P$ is the hydrostatic pressure of the fluid, resulting from the constant volume condition.

The equilibrium structure of the twisted fiber is obtained by minimising the same free energy on a two-dimensional cross-section, exploiting the helical symmetry of the fiber. The cross-section is parametrised in polar form as $(x,y) = \bigl[r_\text{min}+(r_\text{max}-r_\text{min})\,|\!\sin\varphi|\bigr]\bigl(\cos\varphi,\sin\varphi\bigr)$, producing an approximation of the two-cusp nephroid-like cross-section observed experimentally. Along the fiber axis the structure is invariant under a helical symmetry with twist wavevector $k_{\mathrm{tw}}=\pi/L$, where $L$ is the fiber periodicity (a $\pi$-twist along the fiber). The $z$-derivative of the $Q$-field is therefore replaced by an in-plane operator, $\partial_{z}Q_{ij} = k_{\mathrm{tw}}\,[\Omega_{z},Q]_{ij} + k_{\mathrm{tw}}\bigl(y\,\partial_{x}-x\,\partial_{y}\bigr)Q_{ij}$, with $\Omega_{z}=\partial_{z}R\,R^{-1}$ the longitudinal rotation rate, and $R$ the rotation matrix with respect to the reference position. The free energy is integrated over the cross-section. Strong homeotropic anchoring is imposed on the fiber surface in the uniaxial approximation, with the director on the boundary grid points taken consistent with helical symmetry and held fixed. The free energy is minimized in Python with \texttt{scipy.optimize.minimize} (L-BFGS-B), using an analytic Jacobian compiled with Numba.

The cross-section is resolved on an $N\times N$ equidistant mesh with $N=500$, with $\xi/\Delta\approx 2$, $p/\Delta\approx 330$, and $2L/\Delta\approx 680$, where $\Delta$ is the mesh spacing. Geometry and pitch are fixed at values chosen to match the experimental fiber: cross-section diameter ratios $L/d_{\text{min}}=1.08$ and $L/d_{\text{max}}=0.85$ (where $d_{\text{min}}$ and $d_{\text{max}}$ are the short and long diameters of the two-cusp cross-section), and $L/p=1.02$.

To obtain the data in Fig.~\ref{fig:division}, we have written a Python script based on OpenCV to estimate the radii of spherical droplets before and after divisions at different chiral pitches. For each division event, experimentally obtained microscopy frames were selected, and each droplet interface was fitted to a circle. The fitted radius was then used to estimate the volume of each parent and daughter droplet, assuming spherical geometry. To avoid errors in automatically identifying the correct parent-to-daughter droplets and potential mismatches with nearby droplets, we manually assigned an identifier to each parent and daughter droplet. This allowed each droplet to be tracked throughout its full division sequence. The corresponding droplet-tracking sequence is shown in Movie S2. For all scalebars and both automatic and manual size readouts, the microscope magnification and the calibrated camera pixel size were used to convert each pixel reading into micrometers.

\section*{Acknowledgements}
We thank Lindsey Marshall and Boštjan Kokot for VR data exploration using SyGlass, Laurent Weisgerber from DuPont Teijin Films for Mylar spacers, Christian Bahr for fruitful discussions during the preliminary stage of the work, and Natan Osterman for microscope support. VSRJ is thankful to Andrew Fire for encouragement and helpful input during the STS Forum 2022. We also acknowledge funding from the Slovenian Research and Innovation Agency (ARIS) through grants P1-0099 (SČ, MVMR, VSRJ, SZ, MS, SRS, and MR), J1-60002, J1-50006, J1-9149 (SČ), and N1-0400 (VSRJ).

\bibliography{bibliography}

\end{document}